%% file: main.tex
\documentclass[twocolumn,prb,superscriptaddress,floatfix,aps]{revtex4-2}
\pdfoutput=1

\usepackage[utf8]{inputenc}
\usepackage[caption=false,subrefformat=parens,labelformat=parens]{subfig}
\usepackage{graphicx}
\usepackage{float}
\usepackage{amsmath}
\usepackage{amssymb}
\usepackage{dcolumn}
\usepackage{bm}
\usepackage{xcolor}
\usepackage{xfrac}
\usepackage[colorlinks=true, allcolors=blue]{hyperref}
\usepackage{chemformula}
\usepackage[normalem]{ulem}
\usepackage{siunitx}
\usepackage{multirow}

\DeclareSIUnit\angstrom{\text {Å}}
\def\red{\color{red}}

\def\Rtbm{$R\overline{3}m$}
\def\Psmmm{$P6/mmm$}
\newcommand{\Tc}{T_{\mathrm{c}}}
\definecolor{mag}{RGB}{255,0,255}

\begin{document}

\title{{Nb-substitution suppresses the superconducting critical temperature\\of pressurized MoB$_2$}}
\input{authors}
\date{\today}

\begin{abstract}
A recent work has demonstrated that MoB$_2$, transforming to the same structure as \ch{MgB2} ($P6/mmm$), superconducts at temperatures above 30 K near 100 GPa~[C. Pei {\it et al.} Natl. Sci. Rev., nwad034 (2023)], and Nb-substitution in \ch{MoB2} stabilizes the $P6/mmm$ structure down to ambient pressure~[A. C. Hire {\it et al.} Phys. Rev. B 106, 174515 (2022)].
The current work explores the high pressure superconducting behavior of Nb-substituted \ch{MoB2} (Nb$_{0.25}$Mo$_{0.75}$B$_2$).
High pressure x-ray diffraction measurements show that the sample remains in the ambient pressure $P6/mmm$ structure to at least 160 GPa.
Electrical resistivity measurements demonstrate that from an ambient pressure $\Tc$ of 8~K (confirmed by specific heat to be a bulk effect), the critical temperature is suppressed to 4~K at 50 GPa, before gradually rising to 5.5~K at 170 GPa.
The critical temperature at high pressure is thus significantly lower than that found in \ch{MoB2} under pressure (30 K), revealing that Nb-substitution results in a strong suppression of the superconducting critical temperature.
Our calculations indeed find a reduced electron-phonon coupling in Nb$_{0.25}$Mo$_{0.75}$B$_2$, but do not account fully for the observed suppression, which may also arise from inhomogeneity and enhanced spin fluctuations.
\end{abstract}

\maketitle

\section{Introduction} 
The discovery of superconductivity at a critical temperature $\Tc = \SI{39}{K}$ in MgB$_2$~\cite{Nagamatsu2001} two decades ago sparked great interest in diborides amongst the scientific community.
The superconductivity in this material is widely believed to be conventional in nature, {\it i.e.}, deriving from the electron-phonon interaction.
The high critical temperature has been attributed at least partly to high phonon energy scales related to the presence of low mass (light) elements and to the significant covalent character of the states near the Fermi surface~\cite{AnPickett2001,Mazin2003}.
 
A great deal of effort was focused on 
increasing the $\Tc$ to higher values by chemical substitution or pressure.
These attempts were unsuccessful.
Pressure causes a monotonic decrease in the $\Tc$ of \ch{MgB2}~\cite{tomita_dependence_2001, deemyad_dependence_2003}.
Similarly, partial substitutions on the Mg or B sites invariably cause a reduction of $\Tc$~\cite{buzea_review_2001,budko_superconductivity_2015}.
A number of structurally similar borides or borocarbides were also investigated, but none of these exhibited $\Tc$ values comparable to those found in \ch{MgB2}.
A gradual decrease in further exploration of diboride superconductors followed.
On the other hand, the search for high superconducting critical temperatures in light element compounds has been recommenced following the discovery of remarkably high $\Tc$ values in pressurized hydrides~\cite{h3sexp,Somayazulu,Osmond2022}.

The recent discovery of superconductivity in MoB$_2$ with a $\Tc$ reaching as high as \SI{32}{K} at \SI{110}{GPa} has renewed the interest in diborides~\cite{MoB2_superconductivity}.
However, it has been suggested that the mechanism of high $\Tc$ in \ch{MoB2} is significantly different than that in \ch{MgB2}~\cite{Quan2021}.
At ambient pressure \ch{MoB2} exists in an \Rtbm\ structure, which is non-superconducting at low pressure.
Above \SI{25}{GPa}, however, superconductivity appears, with the highest $\Tc$ achieved in the \Psmmm\ phase (the same structure as \ch{MgB2}) at \SI{110}{GPa}.
These results led us to examine whether other diborides might also exhibit remarkably high critical temperatures at elevated pressures.
In a recent paper~\cite{Lim_WB2_2021}, we reported that \ch{WB2} reaches a maximum $\Tc$ of $\sim$$\SI{17}{K}$ at pressures near \SI{90}{GPa}.
Unlike \ch{MoB2}, bulk \ch{WB2} adopts a $P6_3/mmc$ structure over the entire measured pressure range to at least \SI{145}{GPa}.
Our findings suggested that the superconducting nature of \ch{WB2} derives from stacking faults in a \ch{MgB2}-like structure.

An interesting question is whether the superconducting critical temperature of pressurized \ch{MoB2} can be enhanced through chemical substitution.
Our initial work in this direction has focused on examining the effects of partial Nb substitution on the Mo sites because \ch{NbB2} occurs with \Psmmm{} structure in which \ch{MoB2} superconducts above 30 K near 100 GPa.
Recently, we showed, via density functional theory calculations, that phonon free energy stabilizes the \Psmmm{} structure relative to the \Rtbm{} structure at high temperatures across the Nb$_x$Mo$_{1-x}$B$_2$ series~\cite{Hire_NbxMoxB2_2022}.
We were able to successfully synthesize Nb-substituted \ch{MoB2} in the \Psmmm{} structure at ambient pressure via arc-melting.
The resulting compounds, Nb$_{1-x}$Mo$_x$B$_2$, where $x = 0.1, 0.25, 0.5, 0.75, 0.9$, were superconducting with Nb$_{0.25}$Mo$_{0.75}$B$_2$ having the highest $\Tc$ of \SI{8}{K} in the series. 
Specific heat measurements on the $x=0.25$ sample demonstrate bulk superconductivity and also showed a high upper critical field close to \SI{7}{T}~\cite{Hire_NbxMoxB2_2022}. 
In the present study, we further investigate the superconductivity in Nb-substituted \ch{MoB2} ($x=0.25$) through a combination of high-pressure electrical resistivity and x-ray diffraction measurements to pressures as high as \SI{170}{GPa}.

\section{Methods}
At lower pressures ($<$ \SI{2}{GPa}), we used a piston cylinder cell for resistivity measurements~\cite{VanGennep2014}, with the \ch{Nb_{0.25}Mo_{0.75}B2} sample ($\sim$1.0$\times$1.0$\times$\SI{0.4}{\mm^3}) mounted in the van der Pauw configuration.
A solution of n-pentane:isoamyl alcohol (1:1 ratio) was used as the pressure medium.
Details on the use of the piston cylinder cell can be found in Ref.~\cite{Walker_PCC_1999}.

For higher pressure resistivity measurements, a micron-sized \ch{Nb_{0.25}Mo_{0.75}B2} sample ($\sim$40$\times$40$\times$\SI{20}{\micro\m^3}) was placed in a gas membrane-driven diamond anvil cell (OmniDAC from Almax-EasyLab).
A ruby crystal (\SI{20}{\micro\m} in diameter) was used for pressure calibration~\cite{chijioke_ruby_2005} below \SI{80}{GPa}.
At higher pressures, the pressure was determined using the Raman spectrum of the diamond anvil~\cite{Akahama2006}.
Pressure was measured at $\SI{10}{}$ and $\SI{292}{K}$ during each cooling cycle within an error estimation of 5\%.
Two opposing diamond anvils (type Ia, 1/6-carat, \SI{0.15}{\milli\m} central flats) and a cBN-epoxy, soapstone insulated Re metal gasket were used for the four-probe method (see inset in Fig.~\ref{fig:fig1}).
The diamond anvil cell was then placed inside a customized continuous-flow cryostat (Oxford Instruments).
For each temperature-dependent resistivity measurement, pressure was applied at room temperature.
The sample was then cooled to \SI{1.8}{K} before being warmed back to room temperature at a rate of $\sim$$\SI{0.25}{K/min}$.
The measurements were performed with an excitation current of \SI{0.3}{\milli\ampere}.
Further details of the non-hydrostatic high-pressure resistivity techniques are given in Refs.~\cite{Matsuoka2009,Lim_WB2_2021}.

High pressure x-ray diffraction measurements were performed on a powdered piece of \ch{Nb_{0.25}Mo_{0.75}B2} sample at beamline 16-BM-D at the Advanced Photon Source, Argonne National Laboratory.
The x-ray beam had a wavelength of $\SI{0.41}{\angstrom}$ ($\SI{30}{\kilo\electronvolt}$) in Runs 1 and 2, which was focused to a 3$\times$\SI{4}{\micro\m^2} (FWHM) spot at the sample.
A MAR345 image plate detector calibrated with a \ch{CeO2} standard  was used to record the diffracted intensity with the typical exposure time of 60 to 120 seconds per image.
Neon was used as the pressure medium, and pressure was determined both using an online ruby fluorescence measurement~\cite{chijioke_ruby_2005} up to \SI{40}{GPa} as well as the equation of state of Au grains~\cite{Fei_AU_2007} loaded into the sample chamber up to \SI{162}{GPa} within an error estimation of 2\%. DIOPTAS~\cite{Dioptas2015} software was used to convert the 2D diffraction images to 1D diffraction patterns which were further analyzed by Rietveld~\cite{Rietveld_1969} and Le Bail~\cite{LEBAIL1988} methods using GSAS-II software~\cite{Toby_GSASII_2013}.

To better understand the superconducting properties of \ch{Nb_{0.25}Mo_{0.75}B2} under pressure we calculate the Allen-Dynes $\Tc$ at \SI{100}{GPa}.
The electron-phonon coupling constant, $\lambda$, was calculated from Eliashberg spectral function, $\alpha^2F(\omega)$, obtained using the tetrahedron method as implemented in the density functional theory (DFT) code  Quantum Espresso~\cite{qe1,qe2,qe3}.
We use the Perdew–Burke-Ernzerhof functional for the exchange-correlation energy in the DFT calculations~\cite{PBE}.
The virtual crystal approximation was used with the optimized norm-conserving pseudopotentials~\cite{Hamann2013,Schlipf2015}.
A $k$-point mesh of $20\times20\times20$ and a $q$-point mesh of $4\times4\times4$ was used in the calculations.

\section{Results}
\begin{figure}[b]
    \centering
    \includegraphics[width=\columnwidth]{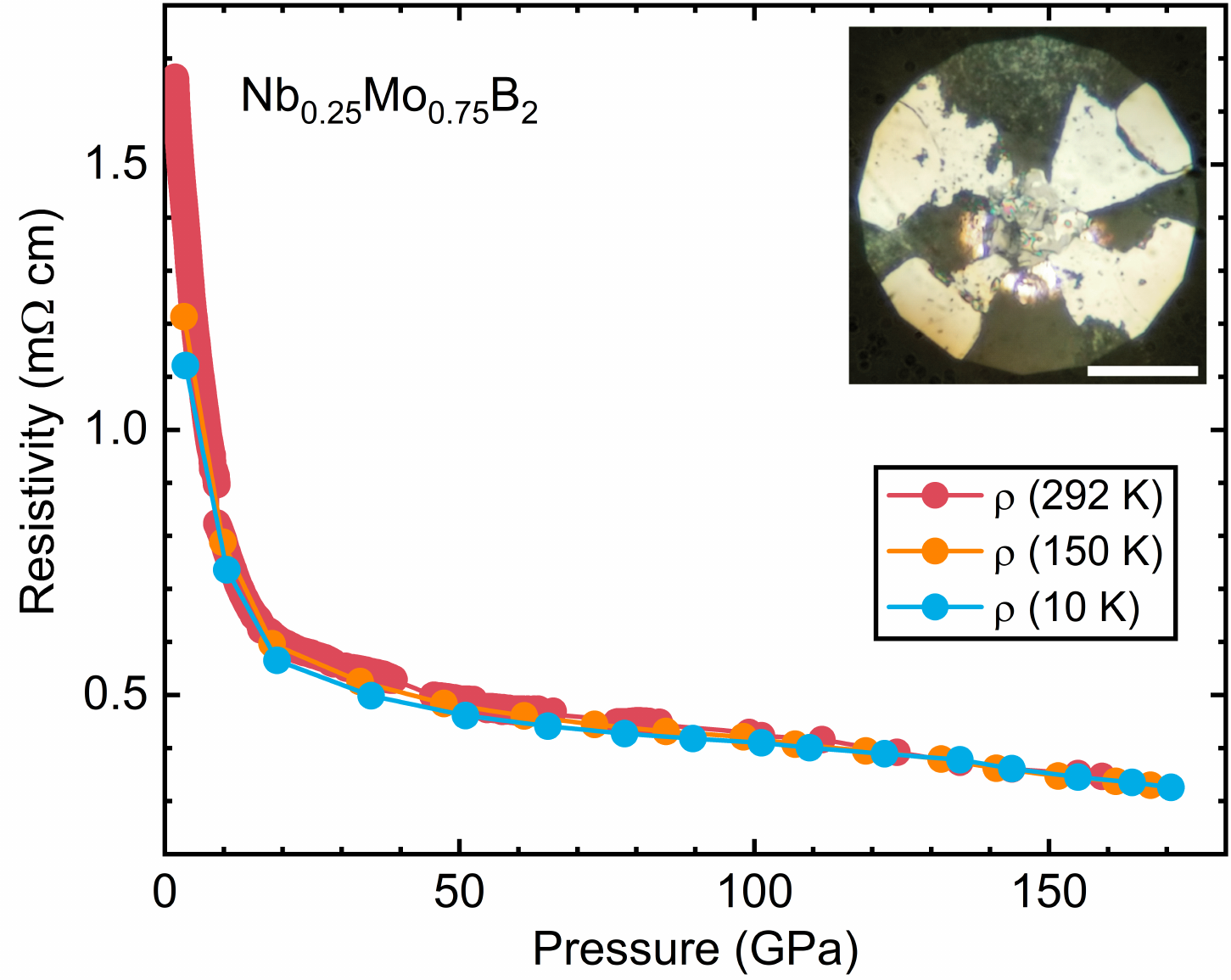}
    \caption{Resistivity of \ch{Nb_{0.25}Mo_{0.75}B2} versus pressure to \SI{171}{GPa} at 10, 150, and \SI{292}{K}. The resistivity curves show no noticeable change with pressure indicating the absence of any structural phase transition. Pressures at \SI{150}{K} were estimated, reflecting the small changes between pressures measured at 10 and \SI{292}{K}. Inset shows the microphotograph of the sample, a ruby pressure calibrant, and the four-probe method looking through the upper diamond central flat (or culet). The white scale bar indicates \SI{50}{\micro\m}.}
    \label{fig:fig1}
\end{figure}
\begin{figure}
    \centering
    \includegraphics[width=\columnwidth]{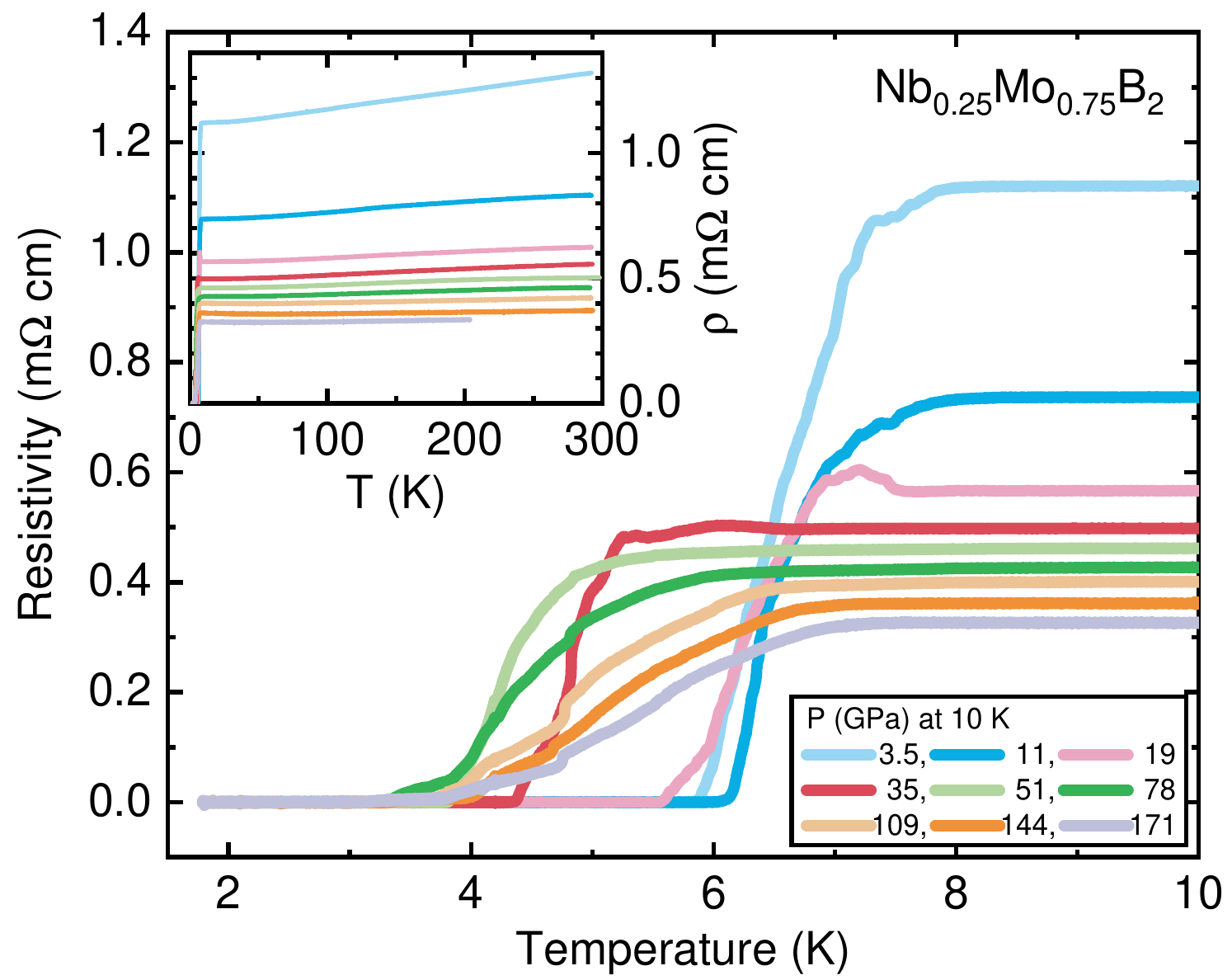}
    \caption{Representative temperature-dependent resistivity curves of \ch{Nb_{0.25}Mo_{0.75}B2} under pressure to \SI{171}{GPa} (measured at \SI{10}{K}) clearly showing the zero resistivity of superconducting transition between 1.8-\SI{10}{K} during each warming cycle. Inset shows the full 1.8-\SI{292}{K} temperature range studied.}
    \label{fig:fig2}
\end{figure}

The pressure-dependent resistivity curves of \ch{Nb_{0.25}Mo_{0.75}B2} are shown in Fig.~\ref{fig:fig1} at 10, 150, and \SI{292}{K}.
While increasing pressure at \SI{292}{K}, the resistivity was measured simultaneously at that temperature.
However, the resistivity curves at 10 and \SI{150}{K} were extracted from the temperature-dependent resistivity at different pressures (see inset in Fig.~\ref{fig:fig2}).
There is no significant change in resistivity with respect to pressure indicating the absence of any structural phase transition.
We also plot the resistivity in a base 10 logarithmic scale showing that the resistivity smoothly decreases with pressure (see Fig.~S1 in the Supplemental Material~\cite{Suppl}).
The inset in Fig.~\ref{fig:fig1} illustrates the four-probe electrical resistivity configuration in the diamond anvil cell looking through the upper diamond used in these measurements.

\begin{figure}
    \centering
    \includegraphics[width=\columnwidth]{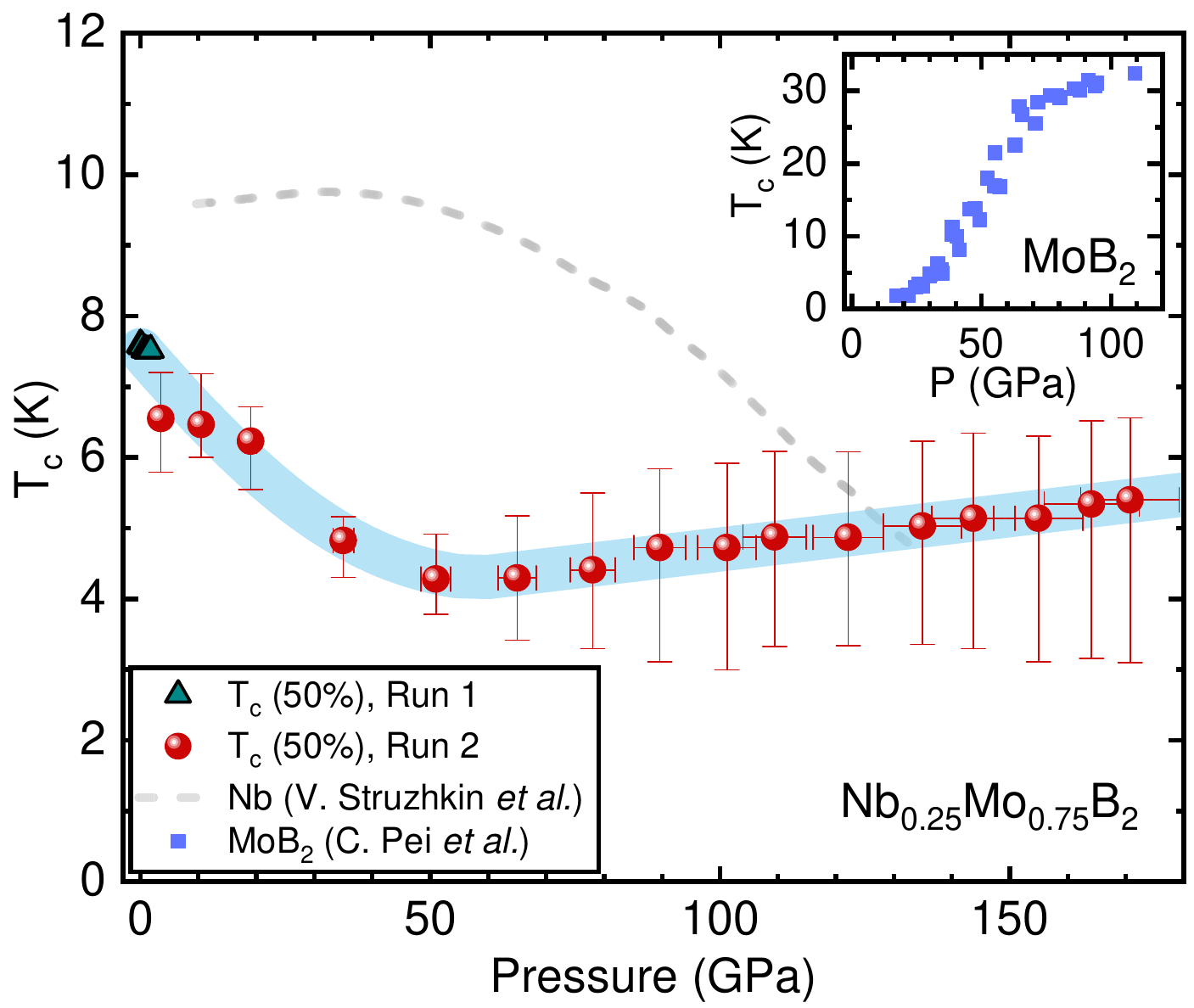}
    \caption{Superconducting phase diagram of \ch{Nb_{0.25}Mo_{0.75}B2} to \SI{171}{GPa} (measured at \SI{10}{K}). The superconducting transition temperature ($\Tc$) initially goes down until $\sim$\SI{50}{GPa} above which it monotonically increases up to \SI{171}{GPa}. The upper and lower vertical bars refer to $\Tc$(90\%) and $\Tc$(offset) respectively. The dashed line shows $\Tc$(P) of elemental \ch{Nb} for comparison~\cite{Struzhkin_TcP_1997}. Inset refers $\Tc$(P) of pure \ch{MoB2} from Ref.~\cite{MoB2_superconductivity}}
    \label{fig:fig3}
\end{figure}

Figure~\ref{fig:fig2} shows selected temperature-dependent resistivity curves under pressures up to \SI{171}{GPa} (measured at \SI{10}{K}) focusing on the superconducting transition.
\ch{Nb_{0.25}Mo_{0.75}B2} superconducts at ambient pressure with a $\Tc$ of \SI{8}{K} as reported by our recent study~\cite{Hire_NbxMoxB2_2022}.
Zero resistivity below the superconducting transition is observed in \ch{Nb_{0.25}Mo_{0.75}B2} throughout the whole pressure range studied.
The superconducting transition broadens significantly above \SI{50}{GPa}. We denote the transition width ($\Delta \Tc$) by vertical bars in Fig.~\ref{fig:fig3}.
The resistivity curve at \SI{171}{GPa} in the inset of Fig.~\ref{fig:fig2} ends at \SI{200}{K}, where the diamonds failed during the warming cycle.
Nevertheless, we managed to measure the highest pressure at \SI{171}{GPa} using diamond anvil Raman at \SI{10}{K} during the cooling cycle (see Fig.~S2 in the Supplemental Material~\cite{Suppl}).

The superconducting transition temperature ($\Tc$) of \ch{Nb_{0.25}Mo_{0.75}B2} versus pressure to \SI{171}{GPa} from Run 1 (below \SI{2}{GPa} including ambient pressure using a piston-cylinder cell) and Run 2 (above \SI{2}{GPa} using a diamond anvil cell) is shown in Fig.~\ref{fig:fig3}.
The $\Tc$(50\%) is defined by the temperature corresponding to the 50\% of normal state resistivity value just above the superconducting transition ($\sim$\SI{10}{K}), whereas the upper and lower vertical bars refer to the 90\% and 0\%(offset) criteria, respectively.
The pressure-dependent superconducting transition temperature ($\Tc$(P)) initially decreases with pressure with a slope of \SI{-0.067(6)}{K/GPa} and above \SI{50}{GPa} monotonically increases with a slope of \SI{0.0097(6)}{K/GPa}.
Interestingly, the slope change in $\Tc$(P) above \SI{50}{GPa} is accompanied by the significant broadening of superconducting transition width ($\Delta \Tc$), defined as the difference between $\Tc$(90\%) and $\Tc$(offset) (see the corresponding vertical bars). The nonhydrostatic condition in the measurement partially contributes to the broadening due to the presence of the pressure gradient. However, the sudden increase above \SI{50}{GPa} suggests the effect originates mainly from the sample itself.
A comparison of $\Tc$(P) between \ch{Nb_{0.25}Mo_{0.75}B2} and elemental \ch{Nb} metal~\cite{Struzhkin_TcP_1997} is shown in Fig~\ref{fig:fig3}, which clearly demonstrates that the superconductivity in \ch{Nb_{0.25}Mo_{0.75}B2} is not associated with Nb inclusions.
Previous work has demonstrated that this material is a bulk superconductor~\cite{Hire_NbxMoxB2_2022}.
\begin{figure}
    \centering
    \includegraphics[width=\columnwidth]{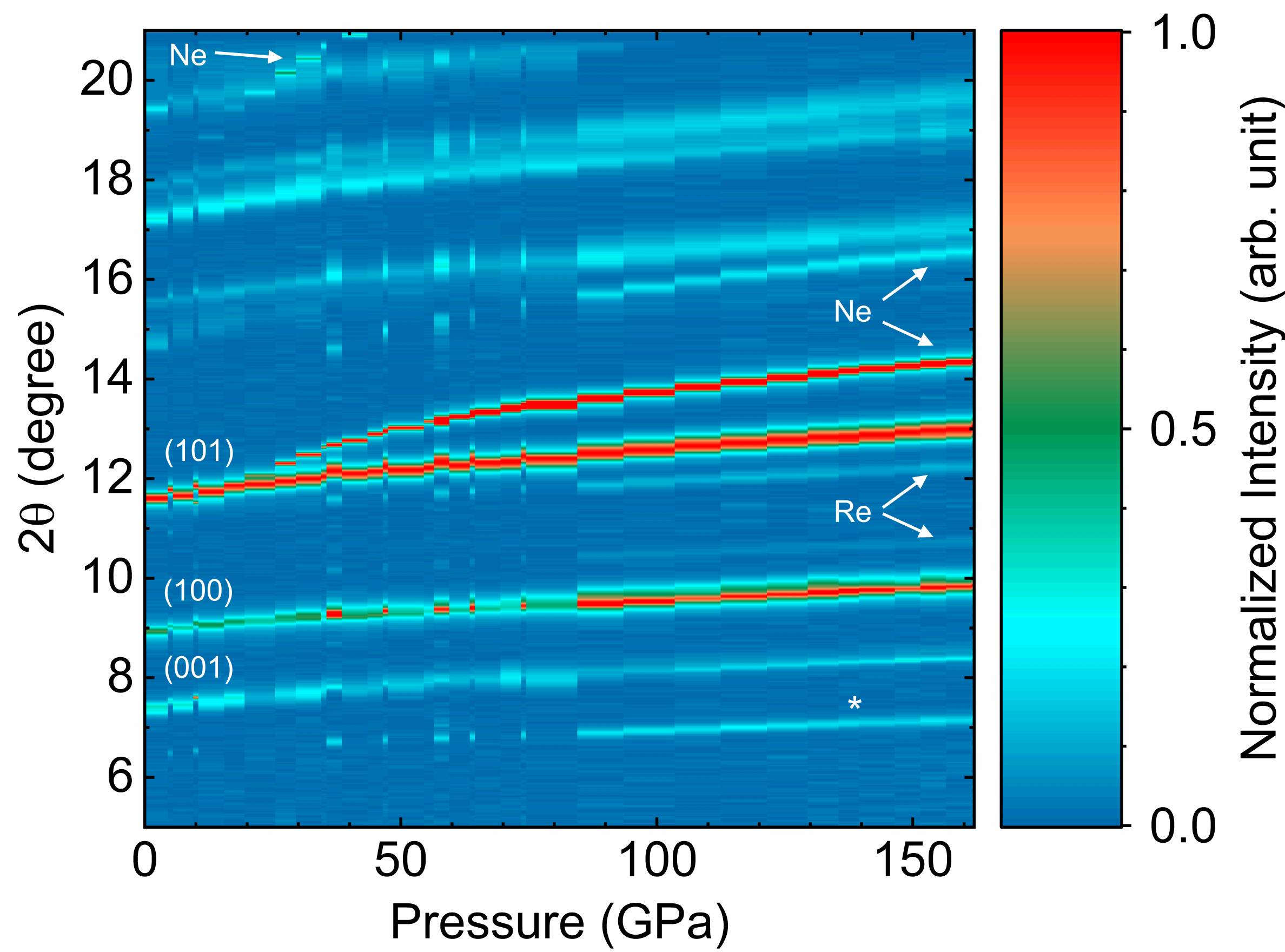}
    \caption{Contour plot of XRD patterns of \ch{Nb_{0.25}Mo_{0.75}B2} to \SI{161}{GPa} at room temperature from Runs 1 and 2. The ambient structure ($P6/mmm$) persists up to the highest pressure without any structural phase transitions.}
    \label{fig:fig4}
\end{figure}

In order to determine the presence of any structural transitions, we have performed synchrotron X-ray diffraction (XRD) measurements on powdered \ch{Nb_{0.25}Mo_{0.75}B2} samples under high pressure and room temperature using Ne as a pressure transmitting medium in diamond anvil cells (DACs).
Figure~\ref{fig:fig4} shows a contour plot of XRD patterns whose intensities are normalized with the (101) peak in Runs 1 and 2.
The $P6/mmm$ structure at ambient pressure persists to pressures as high as \SI{161}{GPa} as seen by the continued presence of the three dominant peaks with (001), (100), and (101) Miller indices.
Vertically offset plots of the XRD patterns with respect to pressure from Runs 1 and 2 are shown in Fig.~S3 in the supplemental material~\cite{Suppl}.
The peaks from the highly compressible Ne can be easily distinguished from those from the sample.
The reflections from both Ne pressure medium and Re metal gasket are confirmed by their equation of state~\cite{Dewaele_NeEOS_2008,Anzellini_ReEOS_2014}.
There is a small amount of unidentified second phase between 6-7 degrees marked by a white asterisk (*).

\begin{figure}
    \centering
    \includegraphics[width=\columnwidth]{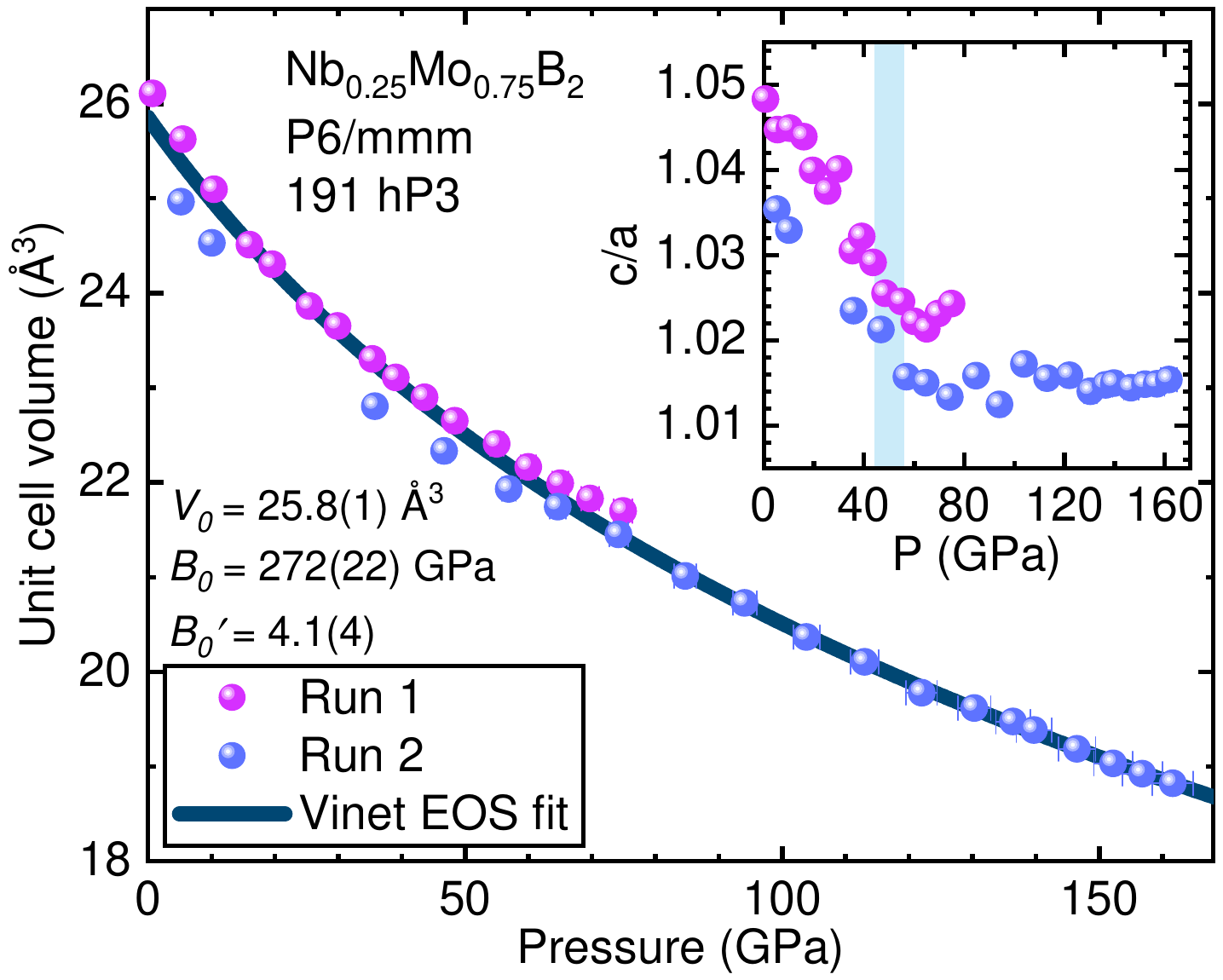}
    \caption{P-V-isotherm of \ch{Nb_{0.25}Mo_{0.75}B2} to \SI{161}{GPa} at room temperature. Inset shows the $c/a$ ratio versus pressure. There is a slope change above $\sim$\SI{50}{GPa} marked by a light blue shaded area referring to the potential correlation with the slope change of $\Tc$(P) in Fig.~\ref{fig:fig3}.}
    \label{fig:fig5}
\end{figure}
The resulting pressure-volume (P-V) curve of \ch{Nb_{0.25}Mo_{0.75}B2} in $P6/mmm$ structure at room temperature from Runs 1 and 2 is shown in Fig.~\ref{fig:fig5} with the $c/a$ ratio versus pressure in the inset.
There is a slope change in the $c/a$ ratio above \SI{50}{GPa} marked by a light blue shaded area, which seems to potentially correlate with the slope change in the $\Tc$(P) in Fig.~\ref{fig:fig3}.
Interestingly, the value of the $c/a$ ratio plateaus above \SI{50}{GPa}, meaning that $c$ lattice parameter begins to be less compressible.
This may indicate that the interaction between interlayers begins to play a significant role in the $P6/mmm$ structure.
The calculated $a$ and $c$ lattice parameters with respect to pressure are shown in Fig.~S4~\cite{Suppl}.
The Vinet Equation of state~\cite{Vinet1987} is used to fit the P-V curve, which gives rise to an ambient volume \SI{25.8}{\angstrom^3} ($V_0$), bulk modulus \SI{272}{GPa} ($B_0$), and a derivative of the bulk modulus of 4.1 ($B_0^{\prime}$).
The bulk modulus of \ch{Nb_{0.25}Mo_{0.75}B2} is comparable to that of \ch{MoB2} (\SI{296}{GPa})~\cite{Yin_moduliWB2_2013}.

Table~\ref{tab:theory_tab_1} shows the computed moments of phonon frequencies, the electron-phonon coupling parameter, and the Allen-Dynes $\Tc$ ($\Tc^{\mathrm{AD}}$) for \ch{NbB2} (at 0 and \SI{100}{GPa}), \ch{Nb_{0.25}Mo_{0.75}B2} (at 50 and \SI{100}{GPa}), and \ch{MoB2} (at \SI{100}{GPa}).
According to these calculations, 25\% Nb-substitution results in a moderate (roughly 30\%) suppression of $\Tc$ compared to pure \ch{MoB2} at \SI{100}{GPa}. This occurs primarily due to a suppression of the electron-phonon coupling.
Interestingly, the calculated $\Tc^{\mathrm{AD}}$ for \ch{Nb_{0.25}Mo_{0.75}B2} at both 50 and \SI{100}{GPa} appear to be overestimations when compared to the experimental $\Tc$. Contrary to the observed experimental trend, we found that $\Tc^{\mathrm{AD}}$ decreases as the pressure increases.
Note that our x-ray diffraction results indicate that at \SI{100}{GPa}, \ch{Nb_{0.25}Mo_{0.75}B2} and \ch{MoB2} adopt the same $P6/mmm$ structure.
\begin{table}
    \centering
    \caption{
        Calculated superconducting parameters.
        The critical temperatures, $\Tc$ were calculated using the Allen-Dynes equation with $\mu^*=0.16$.
        All the calculations utilized the $P6/mmm$ structure.
        The DOS at the Fermi level is in  units of states/eV/unit cell volume.
        (* indicates calculation was performed using the experimental lattice parameters.)}
    \begin{ruledtabular}
    \begin{tabular}{c|c|c|c|c|c|c}
        Material & P & $N(E_{\mathrm{F}})$ & $\omega_{\text{log}}$ & $\langle \omega^2 \rangle$ & $\lambda$ & $T_{\mathrm{c}}^{\mathrm{AD}}$\\
        & (GPa) & & (K) & (K) & & (K)\\
        \hline
        \ch{NbB2} & 0 & - & 354 & 502.6 & 0.75 & 8.86 \\
        \ch{NbB2} & 100 & 0.795 & 577.1 & 767.4 & 0.48 & 1.65 \\
        \ch{Nb_{0.25}Mo_{0.75}B2} & 50 & 1.16 & 268.8 & 426.5 & 1.41 & 23.33\\
        \ch{Nb_{0.25}Mo_{0.75}B2} & 100 & 0.99 & 362.2 & 542.7 & 1.02 & 20.14 \\
        \ch{Nb_{0.25}Mo_{0.75}B2}* & 100 & 0.90 & 419.8 & 608.3 & 0.94 & 19.58 \\
        \ch{MoB2} & 100 & 1.14 & 283.3 & 452.5 & 1.48 & 29.17\\
    \end{tabular}
    \end{ruledtabular}
    \label{tab:theory_tab_1}
\end{table}

\section{Discussion}

One question that still follows from our experiment is why Nb-doped MoB$_{2}$ 
has a significantly lower transition temperatures than MoB$_{2}$ over the same 
pressure range studied in Ref.~\cite{MoB2_superconductivity}. 
Much of the answer to this question can be gleaned from the literature on 
NbB$_{2}$, MoB$_{2}$, and alloyed transition metal diborides. 
We will focus on those findings which are most relevant for superconductivity, 
starting with the density of states (DOS) near the Fermi level.  
When compared with NbB$_{2}$, MoB$_{2}$ has a higher DOS near the Fermi level (Table~\ref{tab:theory_tab_1})
and a higher fraction of electrons occupying antibonding states~\cite{Vajeeston2001,Muzzy2002}. 
This difference helps to explain why, at ambient/low pressure, 
MoB$_{2}$ is a less stable diboride, preferring the trigonal $R\bar{3}m$ space group symmetry with alternating puckered boron planes instead of the hexagonal $P6/mmm$ structure 
realized by NbB$_{2}$~\cite{Vajeeston2001}. 
In addition, MoB$_{2}$ has a higher isotropic electron-phonon coupling constant than 
NbB$_{2}$~\cite{Oguchi2002,Naidyuk2002,Heid2003,Shein2003,Shein2006}. 
Here, we would like to point out that the calculated electron-phonon coupling for \ch{NbB2} at ambient pressure of $\lambda\sim 0.43$ in Singh~\cite{Singh2003} is a result of poorly converged calculations~\cite{Singh2003_b,Heid2003}, and our calculated value agrees with Heid \emph{et al.}~\cite{Heid2003}.

Another interesting aspect of the present study is that the experimentally realized 
suppression of $\Tc$ is at odds with the $\Tc^{\mathrm{AD}}$ obtained using the Allen-Dynes formula. 
The theory and experiment both qualitatively agree that Nb-substitution reduces the $\Tc$ in MoB$_{2}$ at high pressure (Table~\ref{tab:theory_tab_1}) compared with \ch{MoB2}. 
However, there is significant quantitative disagreement in the magnitude of $\Tc$ between the two results. 
Experimentally, we found that \ch{Nb_{0.25}Mo_{0.75}B2} at 100 GPa exhibits only about 30\% of the $\Tc$ of pure MoB$_{2}$ at the same pressure (Table~\ref{tab:theory_tab_1}). 
In contrast, the Allen-Dynes equation predicts that the Nb-substituted sample should exhibit about 70\% of the $\Tc$ of pure MoB$_{2}$ (i.e., for \ch{Nb_{0.25}Mo_{0.75}B2}, $\Tc^{\mathrm{AD}} = 19.58$ - 20.14 K; for \ch{MoB2} $\Tc^{\mathrm{AD}} = 29.17$ K ). 
In other words, the Allen-Dynes $\Tc^{\mathrm{AD}}$ prediction works reasonably well for pure \ch{MoB2}, but it fails to capture the strong reduction in $\Tc$ for Nb-doped \ch{MoB2}.

Performing the same calculation for the $\Tc^{\mathrm{AD}}$ of stoichiometric \ch{NbB2} at ambient pressure reveals a similar overestimation. However, in that case, the degree of overestimation is difficult to gauge since the experimental literature for stoichiometric \ch{NbB2} is rife with inconsistencies.
Some papers report $\Tc$'s between 0.62 K and 9 K~\cite{Leyarovska1979,Schirber1992,Kotegawa2002,Takeya2004}, and many others report an absence of superconductivity down to the lowest temperatures measured~\cite{Cooper1970,Yamamoto2002,Escamilla2006,Geng2007,Regalado2007,Mudgel2008}.
There is considerably more evidence for finite $\Tc$'s up to 8-11 K in \emph{nonstoichiometric} \ch{NbB2}, characterized by increasing the ratio of B to Nb (enabled by Nb-vacancies) ~\cite{Cooper1970,Gasparov2001,Yamamoto2002,Takeya2004,Escamilla2004,Nunes2005,Takahashi2005,Escamilla2006,Regalado2007,Mudgel2008} or decreasing this ratio via B-vacancies~\cite{Hulm1951,Ziegler1953}. Assuming that stoichiometric \ch{NbB2} does not superconduct experimentally, except possibly at minimal temperatures, the Allen-Dynes prediction of $\Tc^{\mathrm{AD}} = $ 8.86 K becomes a  rather severe overestimation. 

In light of the sensitivity to inhomogeneity and vacancy formation in \ch{NbB2}, we point out that \ch{MoB2} is also susceptible to metal vacancy formation, which generally lowers the electronic density of states~\cite{Shein2006}. Taken together, we cannot rule out the role of inhomogeneities due to vacancies in the alloyed sample. Our calculations show that the tendency for metal vacancy formation in Nb$_{0.25}$M$_{0.75}$B$_{2}$ ($E_{\mathrm{vf}}=0.214$ eV) is even more likely than in \ch{NbB2} ($E_{\mathrm{vf}}=1.794$ eV).  The presence of vacancies on the $4d$-atom site could lower the DOS at the Fermi level, reducing $\Tc$. While we do not include these effects in our calculations of the Eliashberg function, we suspect they play a role in the discrepancy between theory and experiment.

Another potential pathology leading to  $\Tc$ predictions larger than experiment could stem from spin fluctuations absent from the present formalism. Several 3$d$ transition metals like V and Cr are better known to have significant spin fluctuations~\cite{Castaing1972,Bauer2014,Tsutsumi2020,Pei2021,Biswas2022}. While Nb is generally considered a conventional electron-phonon superconductor, some claim that spin fluctuations effects are essential for estimating $\Tc$~\cite{Bose_2008,Tsutsumi2020}. We have used a modified McMillan formula defined in Eqn.~(2) of Ref.~\cite{Mazin2002} to estimate the electron-paramagnon coupling constant required to match the experimental $\Tc$ of \ch{Nb_{0.25}Mo_{0.75}B2} (100 GPa), obtaining $\lambda_{\mathrm{sf}}\sim 0.15$. By comparison, to match a $\Tc<0.1$ K in \ch{NbB2} (0 GPa) would require $\lambda_{\mathrm{sf}}> 0.26$. These values are comparable to results for Nb in Ref.~\cite{Bose_2008} and provide at least a partial explanation for the $\Tc$ mismatch. Recent theoretical work on the itinerant antiferromagnet CrB$_{2}$ suggests that spin fluctuations are suppressed under pressure, giving rise to electron-phonon-mediated superconductivity at higher pressures~\cite{Biswas2022}. It is unclear if \ch{Nb_{0.25}Mo_{0.75}B2} exhibits analogous behavior in the pressure dependence of $\Tc$ in part due to the unknown role of other effects like disorder. Further theoretical investigations are necessary to pin down the sources of the overestimation of $\Tc$, which is outside the scope of this study.


Our measured $\Tc$ values are comparable to those reported in many other stoichiometric and 
nonstoichiometric ternary diboride compounds (at ambient/low pressure), such as 
Mo$_{0.95}$Sc$_{0.05}$B$_{2}$ ($\Tc\approx$ 4.8 K)~\cite{Yang2022}, 
Mo$_{0.96}$Zr$_{0.04}$B$_{2}$ ($\Tc\approx$ 5.9 K)~\cite{Muzzy2002}, 
Zr$_{0.96}$V$_{0.04}$B$_{2}$ ($\Tc\approx$ 8.7 K)~\cite{Renesto2013}, 
Zr$_{0.96}$Nb$_{0.04}$B$_{2}$ ($\Tc\approx$ 8.1 K)~\cite{Marques_2016}, 
relevant doped binaries such as Nb$_{1-x}$B$_{2}$ ($\Tc\approx$ 9.2 K)~\cite{Yamamoto2002}, 
NbB$_{x}$ ($\Tc\approx$ 9.4 K)~\cite{Schirber1992}, 
and many other borides of Mo and Nb in the range $\Tc\approx 0$ to 11.2 K~\cite{Cooper1970}. 
There is considerably less literature studying diborides under pressures near 
100 GPa, so it isn't easy to draw complete comparisons with the references 
above. 

In nonstoichiometric \ch{NbB2}, increasing the B/Nb ratio 
tends to expand (shrink) the $c$ ($a$) lattice parameter alongside a concomitant increase in $\Tc$ ~\cite{Cooper1970,Gasparov2001,Yamamoto2002,Takeya2004,Escamilla2004,Nunes2005,Takahashi2005,Escamilla2006, Shein2006, Regalado2007,Mudgel2008}. This behavior indicates that a smaller spacing along the $c$-axis is likely detrimental to superconductivity in \ch{NbB2}. Therefore, one can reasonably expect that the $\Tc$ of NbB$_2$ will decrease under pressure. Our $\Tc$ calculations further support this point, though the actual values are overestimates.
In contrast, experiments by C. Pei {\it et al.} show that the $\Tc$ of MoB$_{2}$ rises sharply with applied pressure beyond 25 GPa until a structural transition near 70 GPa, where $\Tc$ continues to increase with pressure (and the $c$ lattice parameter keeps decreasing) but at a lower rate ~\cite{MoB2_superconductivity}. Hence to achieve a higher $\Tc$ value, both the materials (\ch{NbB2} and \ch{MoB2}) {take advantage of}  different and opposing trends in the lattice parameters. This difference possibly explains the relatively flat $\Tc$ as a function of pressure observed in our experiments. Taken together, we can see that the role of Nb in Nb$_{x}$Mo$_{1-x}$B$_{2}$ is to increase the low-pressure stability of the AlB$_{2}$ structure ($P6/mmm$) without recreating other conditions needed for the higher $\Tc$ observed in MoB$_{2}$ under pressure.


\section{Conclusions}

In summary, we have studied the pressure-dependent superconducting transition temperature of \ch{Nb_{0.25}Mo_{0.75}B2} in the same structure as \ch{MgB2} ($P6/mmm$).
Electrical resistivity measurements up to \SI{171}{GPa} reveal that $\Tc$ initially decreases with increasing pressure.
Above \SI{50}{GPa}, $\Tc$ increases monotonically with a significant broadening of transition width $\Delta \Tc$ up to the highest pressure.
However, the ambient pressure $\Tc$ of $\SI{8}{K}$ is the highest $\Tc$ observed up to at least \SI{171}{GPa}.
Synchrotron high-pressure XRD measurements up to \SI{161}{GPa} show that the slope of the $c/a$ ratio changes above \SI{50}{GPa} within the same $P6/mmm$ structure, indicating a potential correlation with the change in slope of $\Tc$(P).
Our theoretical findings show a reduction of $\Tc$, due to the weakened electron-phonon coupling, in Nb$_{0.25}$Mo$_{0.75}$B$_2$ compared to pure MoB$_2$ at high pressure, in qualitative agreement with the experiment. However, these calculations underestimate the observed suppression of $\Tc$, suggesting that additional factors, such as inhomogeneity and spin fluctuations, may be present.
High-pressure studies of other substitutions into \ch{MoB2}, which might enhance electron-phonon coupling, would be interesting to explore, to determine whether  $\Tc$ values comparable to the \SI{32}{K} observed in \ch{MoB2} at \SI{110}{GPa}~\cite{MoB2_superconductivity} can be realized at low or ambient pressure.

\section*{Acknowledgments}
Work at the University of Florida was performed under the auspices of U.S. Department of Energy Basic Energy Sciences under Contract No.\ DE-SC-0020385 and the U.S. National Science Foundation, Division of Materials Research under Contract No.\ NSF-DMR-2118718.
A.C.H.\ and R.G.H.\ acknowledge additional support from the National Science Foundation under award PHY-1549132 (Center for Bright Beams).
We thank S. Tkachev (GSECARS, University of Chicago) for sample gas loading for the x-ray diffraction measurements, and C. Kenney-Benson (HPCAT) for technical assistance.
R.S.K.\ and R.J.H.\ acknowledge support from the U.S.\ National Science Foundation (DMR-2119308 and DMR-2104881).
X-ray diffraction measurements were performed at HPCAT (Sector 16), Advanced Photon Source (APS), Argonne National Laboratory. HPCAT operations are supported by the DOE-National Nuclear Security Administration (NNSA) Office of Experimental Sciences.
The beamtime was made possible by the Chicago/DOE Alliance Center (CDAC), which is supported by DOE-NNSA (DE-NA0003975).
Use of the gas loading system was supported by COMPRES under NSF Cooperative Agreement No. EAR-1606856 and by GSECARS through NSF grant EAR-1634415 and DOE grant DE-FG02-94ER14466.
The Advanced Photon Source is a DOE Office of Science User Facility operated for the DOE Office of Science by Argonne National Laboratory under Contract No. DE-AC02-06CH11357.
High pressure equipment development at the University of Florida was supported by National Science Foundation CAREER award DMR-1453752.

\bibliography{references}
\end{document}


\title{Supplemental Material: Nb-substitution suppresses the superconducting critical temperature of pressurized MoB$_2$}

\input{authors}

\maketitle

\section{Electrical resistivity}
\begin{figure}[!ht]
  \includegraphics[width=0.7\columnwidth]{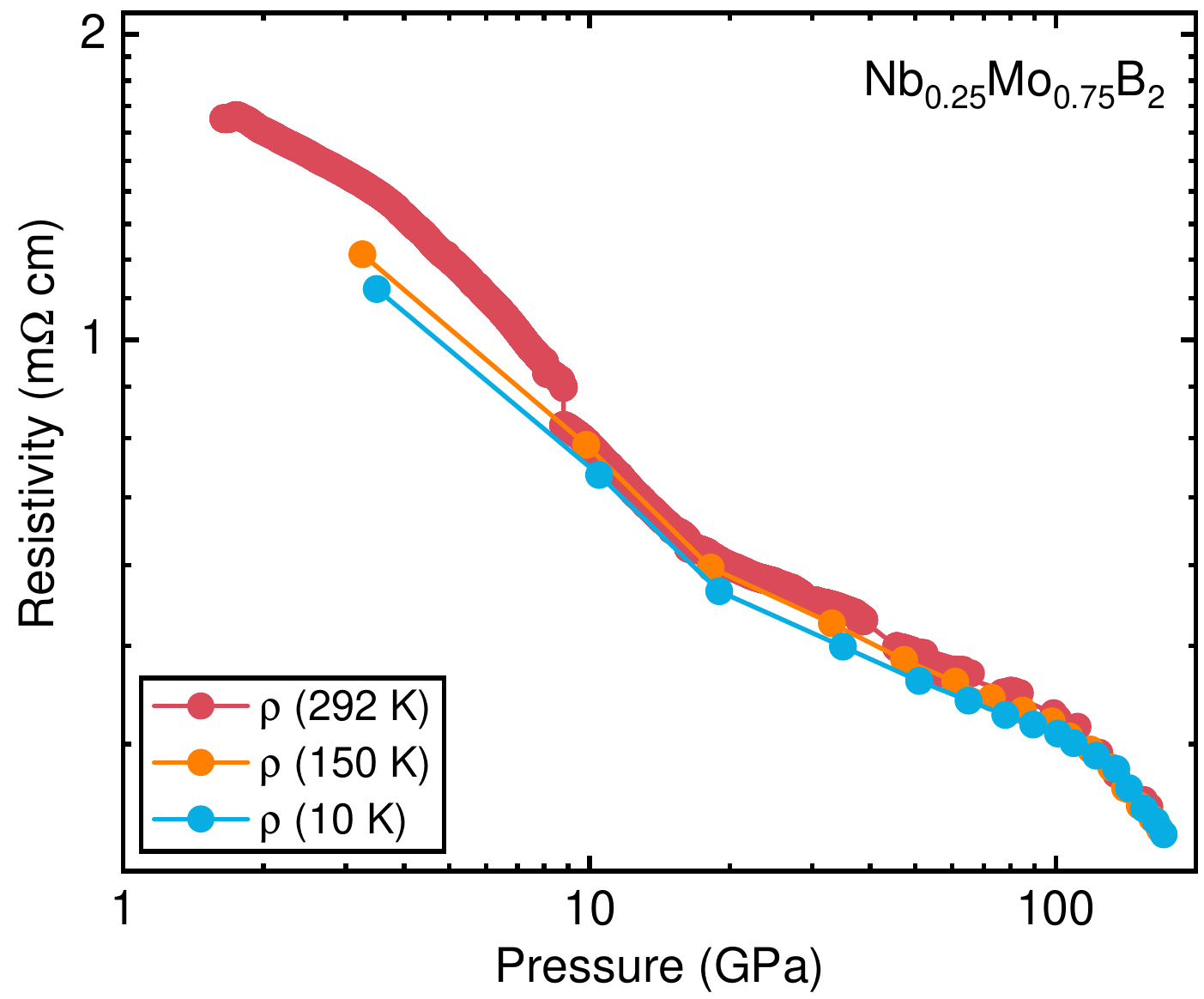}
  \caption{Resistivity of \ch{Nb_{0.25}Mo_{0.75}B2} versus pressure to \SI{171}{GPa} in a base 10 logarithmic scale.}
  \label{FigS1}
\end{figure}

Figure~\ref{FigS1} shows the pressure-dependent resistivity of \ch{Nb_{0.25}Mo_{0.75}B2} to \SI{171}{GPa} in a base 10 logarithmic scale at 10, 150, and \SI{292}{K}. There is no significant change of resistivity with pressure indicating the absence of structural phase transition.

\clearpage
\newpage

\section{Pressure determination}
\begin{figure}[!ht]
  \includegraphics[width=\columnwidth]{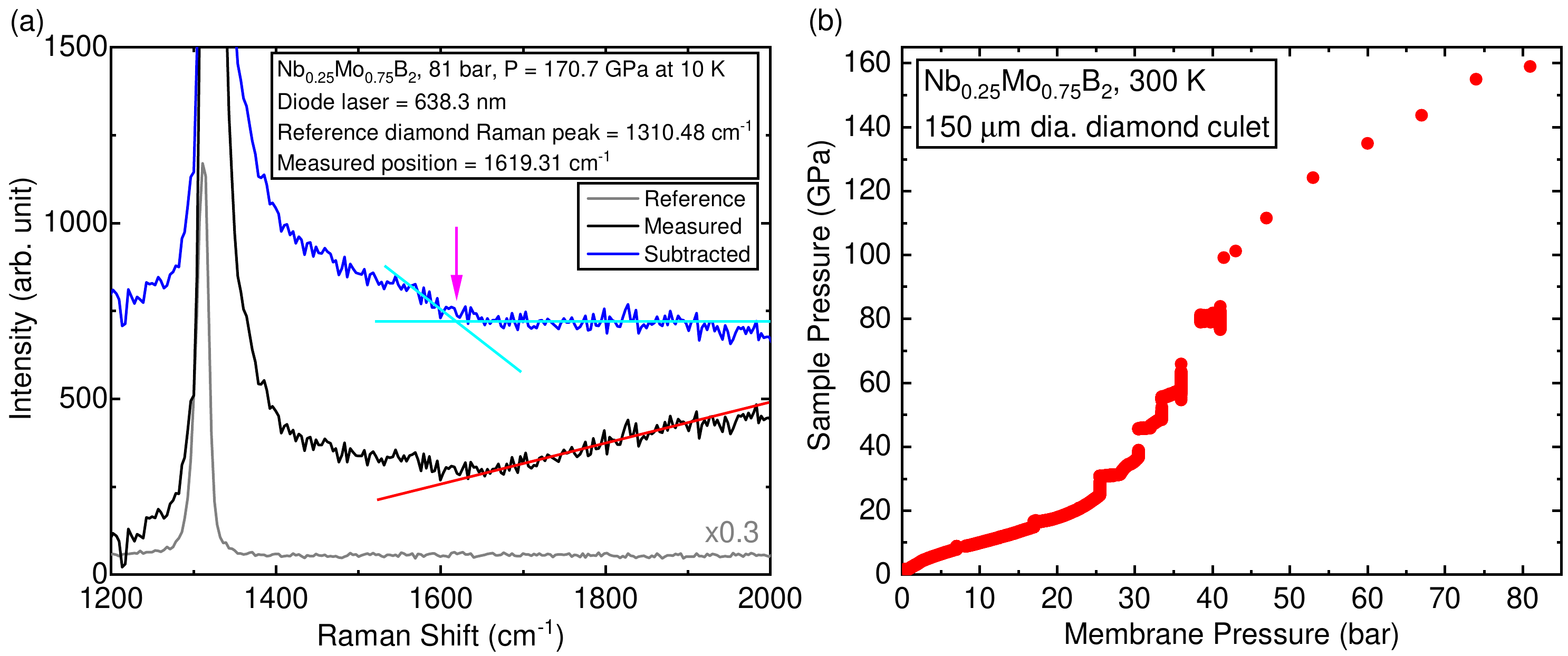}
  \caption{Pressure measurements of (a) diamond anvil Raman and (b) sample pressure versus membrane pressure.}
  \label{FigS2}
\end{figure}

Figure~\ref{FigS2} shows the representative pressure measurements with (a) diamond anvil Raman at \SI{171}{GPa} at \SI{10}{K} and (b) membrane pressure versus sample pressure at \SI{300}{K}. The sample pressure is measured by ruby fluorescence up to $\sim$\SI{80}{GPa}, which can be monitored continuously with increasing membrane pressure as seen in Fig.~\ref{FigS2}(b). Above $\sim$\SI{80}{GPa}, diamond anvil Raman is used at each desired membrane pressure.

\clearpage
\newpage

\section{X-ray diffraction patterns}
\begin{figure}[!ht]
  \includegraphics[width=\columnwidth]{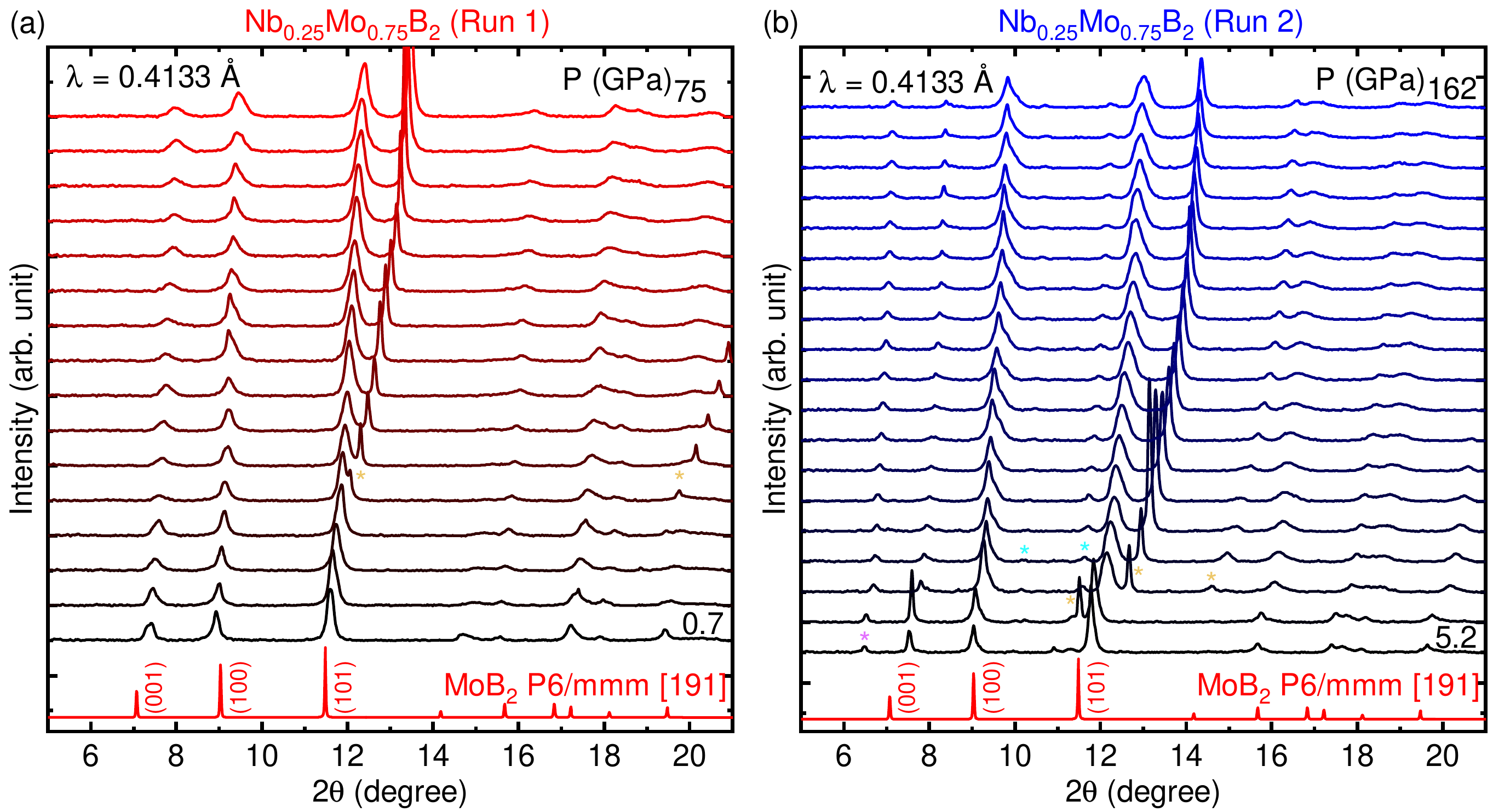}
  \caption{Vertically offset high-pressure XRD patterns of \ch{Nb_{0.25}Mo_{0.75}B2} in (a) Run 1 up to \SI{75}{GPa} and (b) Run 2 up to \SI{162}{GPa}.}
  \label{FigS3}
\end{figure}

Figure~\ref{FigS3} shows XRD patterns of \ch{Nb_{0.25}Mo_{0.75}B2} up to \SI{162}{GPa} at \SI{300}{K}. The ambient structure ($P6/mmm$, 191) persists to the highest pressure studied. No significant discontinuity points to the absence of any structural transition. The light brown * refers to reflections from the Ne pressure medium, whereas the cyan * refers to reflections from the Re metal gasket. The pink * refers to an unidentified second phase with a minor quantity only appearing in Run 2.

\clearpage
\newpage

\section{Calculated $a$ and $c$ lattice parameters up to $\SI{162}{GPa}$}
\begin{figure}[!ht]
  \includegraphics[width=0.7\columnwidth]{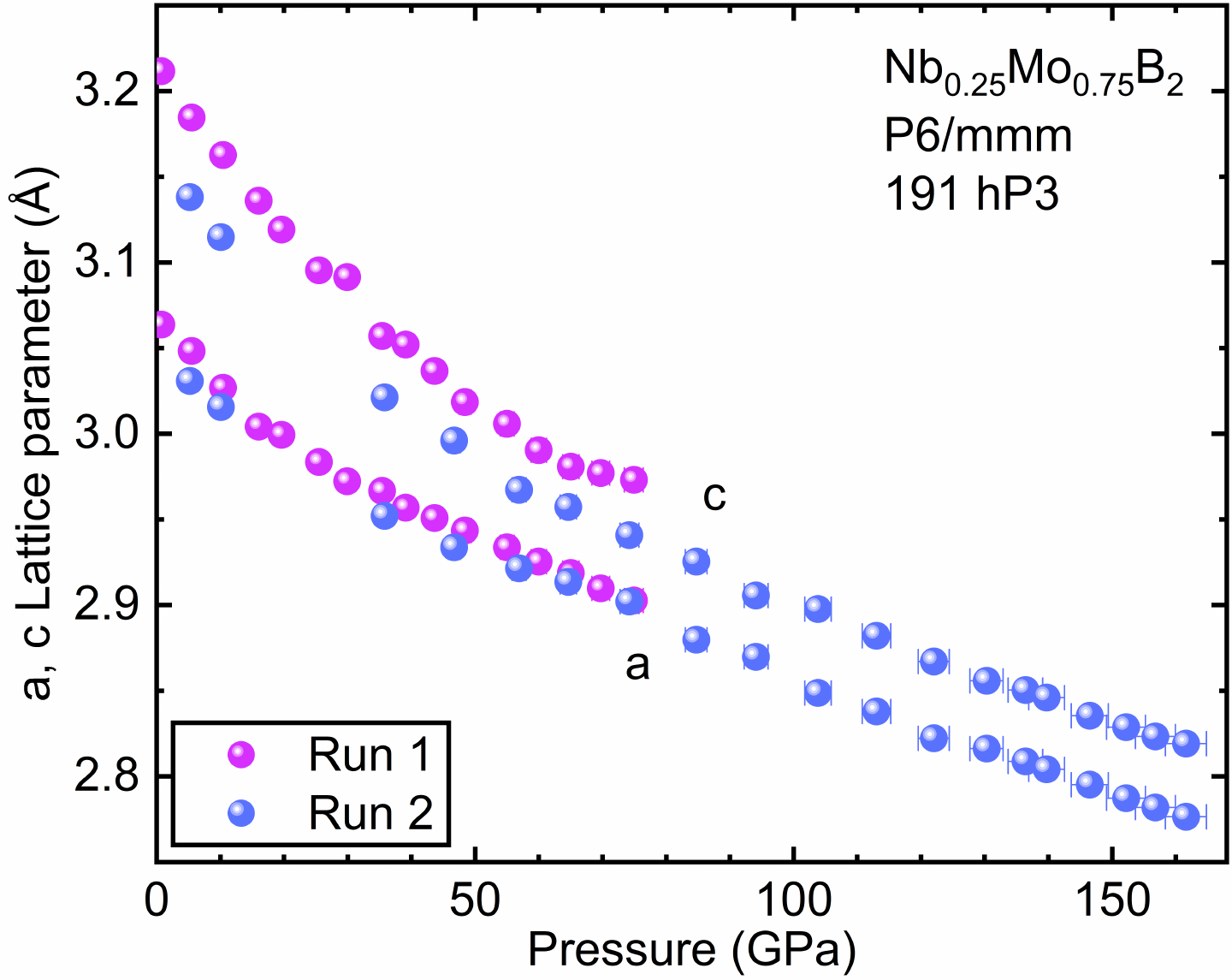}
  \caption{Calculated $a$ and $c$ lattice parameters of \ch{Nb_{0.25}Mo_{0.75}B2} in Runs 1 and 2 up to \SI{162}{GPa}.}
  \label{FigS4}
\end{figure}

Figure~\ref{FigS4} shows a and c lattice parameters of \ch{Nb_{0.25}Mo_{0.75}B2} calculated from the XRD patterns in Fig.~\ref{FigS3} using the Rietveld method with the structure $P6/mmm$ (191).

\clearpage
\newpage

%% file: authors.tex
\author{J. Lim}
\affiliation{Department of Physics, University of Florida, Gainesville, Florida 32611, USA}
\author{S. Sinha}
\affiliation{Department of Physics, University of Florida, Gainesville, Florida 32611, USA}
\author{A. C. Hire}
\affiliation{Department of Materials Science and  Engineering, University of Florida, Gainesville, Florida 32611, USA}
\affiliation{Quantum Theory Project, University of Florida, Gainesville, Florida 32611, USA}
\author{J. S. Kim}
\affiliation{Department of Physics, University of Florida, Gainesville, Florida 32611, USA}
\author{P. M. Dee}
\affiliation{Department of Physics, University of Florida, Gainesville, Florida 32611, USA}
\author{R.\ S.\ Kumar}
\affiliation{Department of Physics, University of Illinois Chicago, Chicago, Illinois 60607, USA}
\author{D.\ Popov}
\affiliation{HPCAT, X-ray Science Division, Argonne National Laboratory, Argonne, Illinois 60439, USA}
\author{R.~J.~Hemley}
\affiliation{Departments of Physics, Chemistry, and Earth and Environmental Sciences, University of Illinois Chicago, Chicago, Illinois 60607, USA}
\author{R. G. Hennig}
\affiliation{Department of Materials Science and  Engineering, University of Florida, Gainesville, Florida 32611, USA}
\affiliation{Quantum Theory Project, University of Florida, Gainesville, Florida 32611, USA}
\author{P. J. Hirschfeld}
\affiliation{Department of Physics, University of Florida, Gainesville, Florida 32611, USA}
\author{G. R. Stewart}
\affiliation{Department of Physics, University of Florida, Gainesville, Florida 32611, USA}
\author{J. J. Hamlin}
\affiliation{Department of Physics, University of Florida, Gainesville, Florida 32611, USA}